# Effects of component-subscription network topology on large-scale data centre performance scaling

Ilango Sriram & Dave Cliff
Department of Computer Science
University of Bristol
Bristol, UK
{ilango, dc} @cs.bris.ac.uk

*Abstract*— Modern large-scale date centres, such as those used for cloud computing service provision, are becoming ever-larger as the operators of those data centres seek to maximise the benefits from economies of scale. With these increases in size comes a growth in system complexity, which is usually problematic. There is an increased desire for automated "self-star" configuration, management, and failure-recovery of the data-centre infrastructure, but many traditional techniques scale much worse than linearly as the number of nodes to be managed increases. As the number of nodes in a median-sized data-centre looks set to increase by two or three orders of magnitude in coming decades, it seems reasonable to attempt to explore and understand the scaling properties of the data-centre middleware *before* such data-centres are constructed. In [1] we presented SPECI, a simulator that predicts aspects of large-scale data-centre middleware performance, concentrating on the influence of status changes such as policy updates or routine node failures. The initial version of SPECI was based on the assumption (taken from our industrial sponsor, a major data-centre provider) that within the data-centre there will be components that work together and need to know the status of other components via "subscriptions" to status-updates from those components. In [1] we used a first-approximation assumption that such subscriptions are distributed wholly at random across the data centre. In this present paper, we explore the effects of introducing more realistic constraints to the structure of the internal network of subscriptions. We contrast the original results from SPECI with new results from simulations exploring the effects of making the data-centre's subscription network have a regular lattice-like structure, and also semi-random network structures resulting from parameterised network generation functions that create "small-world" and "scale-free" networks. We show that for distributed middleware topologies, the structure and distribution of tasks carried out in the data centre can significantly influence the performance overhead imposed by the middleware.

## I. Introduction

Modern large-scale data centres, such as those used for providing cloud computing services, are becoming ever-larger as the operators of those data-centres seek to maximise the benefits from economies of scale. With these increases in size comes a growth in system complexity, which is usually problematic. The growth in complexity manifests itself in two ways. The first is that many conventional management techniques (such as those required for resource-allocation and load-balancing) that work well when controlling a relatively small number of data-centre nodes (a few hundred, say) scale much worse than linearly and hence become impracticable and unworkable when the number of nodes under control are increased by two or three orders of magnitude. The second is that the very large number of individual independent hardware components in modern data centres means that, even with very reliable components, at any one time it is reasonable to expect there always to be one or more significant component failures (so-called "*normal failure*"): guaranteed levels of performance and dependability must be maintained despite this normal failure; and the constancy of normal failure in any one data-centre soon leads to situations where the data-centre has a heterogeneous composition (because exact replacements for failed components cannot always be found) and where that heterogeneous composition is itself constantly changing.

For these reasons, the setup and ongoing management of current and future data-centres clearly presents a number of problems that can properly be considered as issues in the engineering of complex computer systems. For an extended discussion of the issues that arise in the design of warehouse-scale data-centres, see [2].

In almost all of current engineering practice, predictive computer simulations are used to evaluate possible designs before they go into production. Simulation studies allow for the rapid exploration and evaluation of design alternatives, and can help to avoid costly mistakes. Computational Fluid Dynamics (CFD) simulations are routinely used to understand and refine the aerodynamics of designs for airplanes, ground vehicles, and structures such as buildings and bridges; and to understand and refine the hydrodynamics of water-vehicles. In microelectronics, the well-known SPICE circuit-simulation system [3] has long allowed large-scale, highly complex designs to be evaluated, verified, and validated in simulation before the expensive final stage of physical fabrication.

Despite this well-established tradition of computational modelling and simulation tools such as CFD or SPICE being

Financial support for this work came from the Hewlett-Packard Automated Infrastructure Lab, HP Labs Bristol (for I. Sriram) and the UK Engineering and Physical Sciences Research Council (EPSRC)'s Large-Scale Complex IT Systems (LSCITS) Research Initiative (for D. Cliff).

used in other engineering domains, there are currently no comparable tools for cloud-scale computing data-centres. The lack of such tools prevents the application of rigorous formal methods for testing and verifying designs before they go into production. Put bluntly, at the leading edge of data-centre design and implementation, current practice is much more art than science, and this imprecision can lead to costly errors.

As an exploratory step in meeting this need, we have developed SPECI (Simulation Program for Elastic Cloud Infrastructures). Clearly, it would require many (tens or hundreds of) person-years of effort to bring SPECI up to the comprehensive level of SPICE or of commercial industrial-strength CFD tools. We are currently exploring the possibility of open-sourcing SPECI in the hope that a community of contributors then helps refine and extend it.

The first paper discussing SPECI [1] gave details of its rationale and design architecture that will not be repeated here. In that first paper, results were presented from simulation experiments that had been suggested by our industrial sponsor, Hewlett-Packard Laboratories. The specific area of inquiry in [1], and here also, is large-scale data-centre middleware component-status subscription-update policies.

The status of data-centre components may change as they fail, or as policies are updated. Within the data-centre there will be components that work together and need to know the status of other components via "subscriptions" to status-updates from those components. In [1] we used a first-approximation assumption that such subscriptions are distributed randomly across the data centre. That is, the connectivity of the network of subscription dependencies within the data-centre is, formally, a random graph. In this present paper, we explore the effects of introducing more realistic constraints to the structure of the internal network of subscriptions. We contrast the original results from SPECI with new results from simulations exploring the effects of making the data-centre's subscription network have a regular lattice-like structure, and also the effects when the network has semi-random structures resulting from parameterised network generation functions that create "small-world" [4] and "scale-free" [5] networks. We show that for distributed middleware topology, varying the structure and distribution of tasks carried out in the data centre can significantly influence the performance overhead imposed by the middleware.

In reality, component subscription connectivity will be affected by the physical layout of the data-centre hardware, and by the virtual placement of services in that data-centre. There are multiple physical levels of layout. Subscriptions to components resident on the same piece of silicon as the subscribing component have the highest data-transfer rates and lowest transmission latency, but in general the subscribed components will be resident elsewhere: in another chip on the same server motherboard; or on another motherboard in the same vertical rack of servers; or on another rack in the same unit (aisle or cluster) in the data-centre, or in another unit elsewhere in the same data centre (merely under the same roof); or perhaps in another data-centre tens, hundreds, or thousands of miles away. Network bandwidth is commonly a scarce resource in data-centre management, so communications bandwidths and latencies increase as the locations of the subscribed components become ever more distant. The work reported here demonstrates that SPECI can accommodate varying network topologies; future work is planned to accurately model one or more real-world data-centres.

We see the results in this paper as the first step toward developing adaptive data-centre management policies that can "intelligently" organise and reorganise the network of subscriptions within the data-centre in light of changing demands, and deal robustly with the effects of normal failure.

The structure of this paper is as follows. In Section II we give further details of SPECI, sufficient for the reader to comprehend the new results presented in this paper. In Section III, for completeness, we summarise the results presented in [1] as those results form the baseline against which we then compare the outputs from the more structured subscription networks. Section IV explains the structures of subscriptions used and their implementation, and the results from these simulations is shown in Section V. In Section VI we discuss future work, and close with a conclusion in Section VII.

## II. EXPLANATION OF SPECI

In this section we reiterate segments from [1], which are necessary to give the reader an understanding of the purpose and functionality of SPECI, and are necessary for understanding the output of SPECI that will be discussed in the following sections.

### A. Middleware Scalability & Inconsistencies from Failure

Cloud-scale data centres are built using commodity hardware, and rely on inexpensive traditional server architectures with the key components being CPU time, memory usage, disk space, and network connectivity. As economies of scale are driving the growth of these data centres (DCs), the sheer number of off-the-shelf components used in coming decades, in combination with each component's average life cycle will imply that component failure will occur continually and not just in exceptional or unusual cases. This expected near-permanent failing of components is called "*normal failure*". For cost reasons, the DC operator will leave the failed components in place and from time to time replace the servers on which failure occurred or even entire racks on which several servers have failed. The impact of failure and resilience or recovery needs to be taken into account in the overall performance assessment of the system.

The components of the DCs are tethered by a software layer (so-called "middleware") that is responsible for job scheduling, load-balancing, security, virtual network provisioning, and resilience. It combines the parts of the DC and is the management layer of the DC. As the numbers of components in the DC increases, the middleware has more to handle. Scalability requires the performance not to rapidly degrade as the number of components increases, so that it remains feasible to operate in the desired size range, and also that the system remains deployable economically, productively, and with adequate quality of service [7]. Yet it is unlikely that all properties in middleware will scale linearly when scaling up the size of DCs.

Because the middleware's settings and available resources change very frequently, it needs to continuously communicate new policies to the nodes. Traditionally middleware manages its constituent nodes using central control nodes, but hierarchical designs scale poorly. Distributed systems management suggests controlling large DCs using policies that can be broken into components for distribution via peer-to-peer (P2P) communication channels, and executed locally at each node [8]. P2P solutions scale better, but can suffer from problems of *timeliness* (how quickly updated policies will be available at every node) and of *consistency* (whether the same policies are available and in place everywhere).

Status updates have been successfully utilised to develop Aneka-Federation, a fully decentralised cloud resource broker implemented as an overlay service to coordinate application scheduling and to federate resources from multiple clouds [9]. Aneka-Federation gains scalability and fault-tolerance by using Distributed Hash Tables (DHTs).

Consistency protocols like gossiping (epidemic spread) have also achieved scalable and fault-tolerant information dissemination [10]. However, gossip protocols usually come with long delays in transporting messages, and are found not to be robust towards correlated failure [11]. This makes gossip undesirable as underlying protocol for middleware distribution. In either management form, a certain overhead load for the management will be generated, which will determine the performance loss when scaling up the DC by adding more components.

As a first step, SPECI has been constructed to explore the behaviour of a key part of the middleware: that which recognises failed components across the network of systems. This failure communication mechanism can be seen as a simplified substitution for the policy distribution problem.

### B. Setup

We are interested in the behaviour of systems with a large number of components, where each component can be working correctly or exhibiting a temporary or permanent failure. Any one component cooperates with some of the other components, is thus interested in the aliveness of those other components, and actively performs queries to find this out, e.g. by polling the other components for their "aliveness" state. As the number of components increases, the number of states that have to be communicated over the network increases. We need to know what happens with our system in terms of how well in time can the states be communicated and at the cost of what load. This setup is of interest to any computing facility, which has such a large number of components that the expected number of failures at any one time is significantly above zero, or indeed in large-scale situations where other changes need to be communicated frequently. The key issues examined in this use of SPECI are the scaling properties of various protocols, and how quickly (if at all) a consistent view of the state of cooperating nodes can be achieved under certain conditions. Our simulation methods are described in the following paragraphs.

There is a number ($n$) of nodes or services connected through a network. Each of these nodes can be functioning (alive) or not (dead). To discover the aliveness of other nodes, each node provides an arbitrary state to which other nodes can listen. When the state can be retrieved the node is alive, otherwise it is dead. The retrieval of aliveness of other components is referred to here as the "heartbeat". Every node is interested in the aliveness of some of the other nodes, the value of "some" being a variable across all nodes. Each node maintains a subscription list of nodes in whose aliveness it is interested. We are interested in how the implementation of the heartbeat affects the system under given configurations, when the total number of nodes $n$ increases.

Several heartbeat protocols are possible. We explore four here: First, the *Centralised* approach where central monitoring nodes collect the aliveness of all other nodes, and then inform any node interested in any particular state. Second, the *Hierarchical* approach where, depending on the number of hierarchy levels, certain nodes would gather the information of some other nodes, and make them available to their members and to the node next higher in the hierarchy. Third, a *Simple P2P* mechanism where any node simply contacts the node of interest directly. Fourth, a smarter *Transitive P2P* protocol where a contacted node would automatically reply with *all* the aliveness information it has available for other relevant nodes.

The investigation reported here was set up to observe the behaviour of the overall system under these protocols, for various change-rates, when the number of nodes $n$ is scaled up over several orders of magnitude. The simulations address a number of questions. The first questions of interest are: what is the overall network load for each of the above protocols under given settings and size, and how much data has to be sent over the network in a given time period? Second, there is significant commercial interest in what the "time-for-consistency" curve of the system looks like. That is, after simultaneous failure or recovery of a number of nodes, after how many time-steps are the relevant aliveness changes propagated through the entire system, and if there are continuous failures appearing, how many nodes have a consistent view of the system over time? It is of further interest to see how many time-steps and how much load it takes until new or recovered nodes have a consistent view of the system, and how many time-steps it takes to recover after failure of a large fraction of the $n$ nodes, or for recovery of the entire network. There is also interest in the trade-off between timeliness and load for each of the protocols in the sense of how much extra load will be required to retrieve a better or more consistent view. In other words, for how much load can one get what degree of timeliness?

All runs were carried out for each of the four heartbeat protocols (Centralised, Hierarchical, Simple P2P, & Transitive P2P). SPECI provides a monitoring probe of the current number of inconsistencies, and the number of network packets dealt with by every node, every second. For now, the simulator assumes uniform costs for connecting to other nodes, but we intend to explore varying the connection costs in a meaningful way, in future work.

III. BASELINE RESULTS

In this section, for completeness, we summarise the random-connectivity SPECI results presented in [1], as these results will form the baseline against which we then compare the outputs from more structured topologies of subscription networks. In addition to those results, we will discuss the measurements of load data from experiments with the original setup in this section, as these will be used for comparison in the following sections, too.

Initially, we observed the number of nodes that have an inconsistent view of the system. A node has an inconsistent view if any of the subscriptions that node has contains incorrect aliveness information. We measure this as the number of inconsistent nodes, observed here once per $\Delta t$ (=1sec). After an individual failure or change occurs, there are as many inconsistencies as there are nodes subscribed to the failed node. Some of these will regain a consistent view within $\Delta t$, i.e. before the following observation, and the remaining ones will be counted as inconsistent at this observation point. If the recovery is quicker than the time to the next failure, at the subsequent observations fewer nodes will be inconsistent, until the number drops to zero. When the heartbeat method requires aliveness data to be passed on, more hops would make us expect more inconsistencies, as outdated data could be passed on. This probing was carried out while running SPECI with increasing failure rates and scale, and using each of these combinations with each of our four heartbeat protocols. We scaled $n$ though DC sizes of $10^2$, $10^3$, and $10^4$ nodes (experiments for $n=10^5$ and $n=10^6$ are underway, and we expect to show those results in the final published version of this paper). We assume that the number of subscriptions grows slower than the number of nodes in a DC, and so we set the number of subscriptions to $n^{0.5}$ per node.

For each of these sizes a failure distribution $f$ was chosen such that on average in every minute 0.01%, 0.1%, 1%, and 10% of the nodes would fail. Because this work is essentially exploratory, a gamma distribution was used with coefficients that would result in the desired rate of failures. Each pair of configurations was tested over 10 independent runs, each lasting 3600 simulation time seconds, and the average number of inconsistencies along with its standard deviation, maximum, and minimum number were observed. The half width of the 95% confidence intervals (95% CI) was then calculated using the Student's t-distribution for small or incomplete data sets.

Figure 1 shows the average fraction of nodes whose information about their subscriptions is inconsistent with the real state. The fraction increases when the failure rate increases, and also when the number of nodes increases. This was well expected, and the growth with a higher failure rate seems obvious, though it was not obvious that the rate would also grow with the number of components. Figure 2 shows the same data as Figure 1, but plotted on a linear scale to render the confidence intervals more visibly: it is clear that for small $n$ and small failure rates the heartbeat protocols differed insignificantly. But, as n and the failure rates increase, significant differences between the protocols emerge. In Figure 2 this difference between the central and hierarchical protocols and the two P2P-based protocols can first be seen for $10^3$ nodes and 10% failure rate. In [1] we only illustrated the Hierarchical and TransitiveP2P protocols; there we showed that for $n=10^3$ the differences are significant from 1% onwards. This demonstrated that with increasing node-counts and failure rates, the choice of the protocol becomes significant, and also that the P2P protocol scales better for the objective of low inconsistencies under the initial simplifying assumptions. Given that there were identical polling intervals, the TransitiveP2P was expected to be the protocol with the largest number of inconsistencies. This is due to the fact that the transitiveness in the protocol would allow forwarding of delayed data, and those delays could accumulate and lead to propagation of out-of-date status information. However, due to the random-graph nature of the subscription network, the clustering coefficient in our studies was not sufficient for this effect to be reliably observed. We will come back to explore this effect in Section V. Figure 3 shows the same inconsistency data, but grouped by failure rates. As in all the figures here illustrating inconsistencies, to be able to compare the values of different DC sizes, the number of inconsistencies is normalised by the number of nodes and the fraction of inconsistencies shown. Despite the normalisation, the number of inconsistencies still grows with the size of the DC: none of the protocols scale linearly. On the other hand, when grouped by $n$, these normalised values increase by one order of magnitude when the failure rate increases by one order of magnitude. This linearity suggests that the failure tolerance of all four protocols appears robust. Figure 4 shows the load measured in network access counts during these runs. Under the Centralised protocol, the load grows significantly faster with increasing $n$ than under the other three protocols. In this case, there are only minor scaling differences from one size to the next between the various protocols, so one could imagine that differences are more in the implementation detail than in the structure. However, this load is for identical polling intervals between the protocols. To reduce the inconsistencies discussed earlier, in both centralised protocols the polling frequency needs to be increased at the cost of additional load.

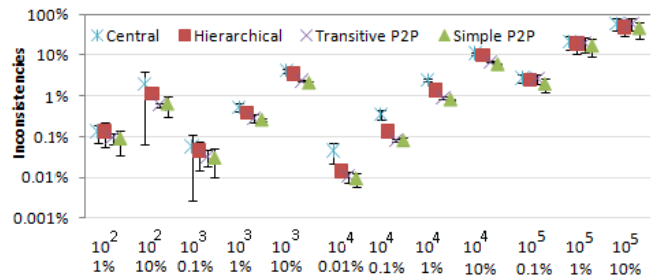

Figure 1. The mean percentage of inconsistencies (vertical axis) increases with the number of nodes and the percentage failure rate (horizontal axis).

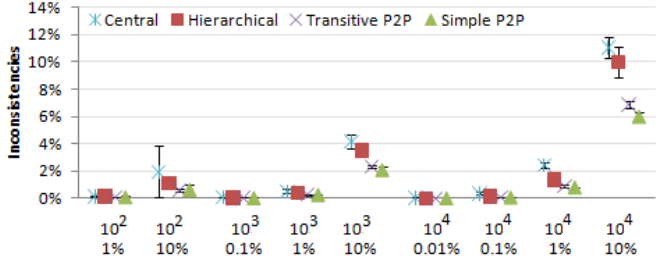

Figure 2. Mean of Inconsistencies and their confidence intervals. In [1] we showed that with increasing size differences between the protocols become significant.

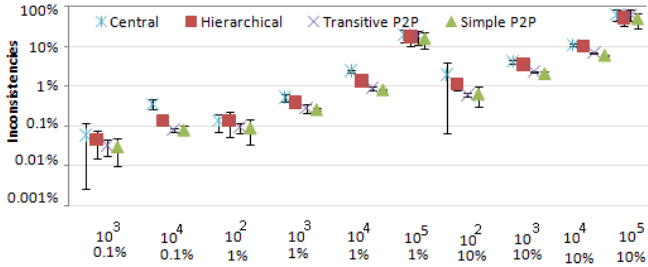

Figure 3. Inconsistencies grouped by failure rate. Although the values are normalised by the scale, there is still an increase. This suggests that none of the protocols scale linearly.

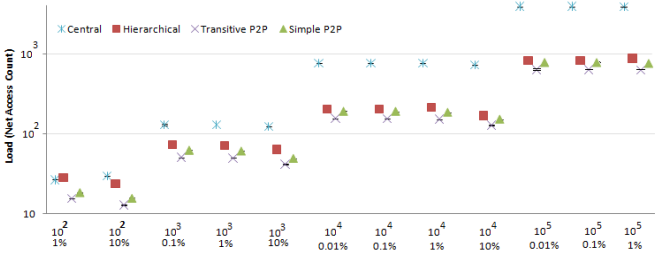

Figure 4. With the setting of regular polling intervals, the average load per node does not vary with the failure rate.

## IV. STRUCTURED DISTRIBUTIONS OF SUBSCRIPTIONS

Section III summarised the outcome of simulations based on a first-approximation assumption that the subscriptions being polled for aliveness are distributed randomly across the data centre. That is, the graph of the network of subscription dependencies within the data-centre is, formally, a random graph. In the following, we explore the effects of introducing more realistic constraints to the structure of the internal network of subscriptions. We contrast the previous results from SPECI with new results from simulations exploring the effects of making the data-centre's subscription network have a regular lattice-like structure, and also the effects when the network has semi-random structures resulting from parameterised network generation functions that create "small-world" and "scale-free" networks. In Subsection A we sketch out the properties of these networks, and in Subsection B we explain the implementation details used. The SPECI results from using these subscription structures are then presented in Section V.

### A. Structures used for subscription networks

**Lattices.** Subscriptions drawn from a lattice or a two-dimensional grid are connected to the close neighbourhood of the node. This is a two dimensional variation of the common k-neighbour graph. Two-dimensional lattices exhibit a high clustering coefficient, which means there is a high probability that two nodes that are subscribed to each other have further common subscriptions. In the real world, such a form of subscriptions could be expected when the closest possible placement of collaborating nodes is chosen without any perturbations.

**Small worlds.** The key property of "small world" networks is that they have the same high clustering coefficient (i.e. many of the neighbours are themselves neighbours) as regular lattice graphs have, but at the same time they have low diameters (i.e. short average path lengths) as found in random graphs. Watts and Strogatz proposed a model in which they merged local and long-range "contacts" and described the algorithm as rewiring a ring lattice [4]. They also showed that networks with such properties arise naturally in many fields and are commonly found in natural phenomena. In a data centre, we could imagine the subscription graph having a small-world distribution, when components are initially placed close to each other, and over time change their location or their functionality (e.g. as a result of load-balancing) thereby turning from local into long range contacts.

**Scale-free.** Scale-free networks are networks with a power-law degree distribution. They gained popularity when Barabási and Albert proposed an algorithm for creating them by growing networks with preferential attachment [5]. Scale-free networks have been found in many complex networks in the real world, too. For instance, most of the properties in the internet are scale-free networks. Scale-free networks could have applicability for subscriptions when there are components with different importance for other nodes, for instance.

For a comprehensive review of complex networks, and also for a discussion of weaknesses of directed versions of the Barabási Albert model see the Newman's review of complex networks [12].

### B. Implementation of subscription networks

#### 1) Nearest neighbours on lattice

To create a lattice-like structure, the nodes are numbered from 0 to *n*-1, and aligned on a grid with the width (or number of columns) set to the largest integer smaller or equal to the square root of the number of nodes. The grid is then filled row by row. Thus, the number of lines in this grid is either the same as the number of columns, when *n* is a square number, or there are a few more rows than columns, and the last row does not have to be filled entirely. Each node in this grid is then subscribed to approximately its *k* nearest neighbours: This is done by starting with the node located above the current one in the grid, and then iterating coordinates clockwise around the already subscribed nodes, until the total number of subscriptions is reached. Periodic boundary conditions are used, so that when the destination coordinate reaches the end of the grid, it jumps to the corresponding position on the opposite side. With this approach, each node will have exactly *k* subscriptions.

*2) Small-world subscriptions*

The original algorithm to generate small-world random networks was proposed by Watts and Strogatz in [4]. Their model is an undirected network that merges properties of local and long-range "contacts" by starting the network from a ring lattice, i.e. a network with local contacts, and changing some edges to turn into long-range contacts, the way they would appear in small-world networks. To implement small-world networks, the first step is to create a ring lattice and connect every node with its *k* nearest neighbours. In the second step, one introduces long-range contacts by rewiring a small number of these edges to point to a new end or node chosen at random. To model the subscriptions of the nodes for the at hand work in this way, a directed version of this algorithm is needed. In his work searching for a decentralised algorithm to find shortest paths in directed small-worlds, Kleinberg [13] has proposed a directed version of small-world networks in which he starts from a two-dimensional grid rather than a ring. However, here we stay closer to the original algorithm by Watts and Strogatz and simply start with directed edges in a ring lattice. Each edge has outgoing edges to its *k* nearest neighbours, *k* being the number of subscriptions in our model. Thus, between neighbouring edges there will be two edges, one in each direction. The second pass is implemented as in the original undirected algorithm. With a low probability *p* each subscription is rewired in such a way that the originating node stays the same, and the new destination vertex is picked uniformly from the nodes to which the current node has no connection to. Here we used *p*=0.1, which in Watts and Strogatz' work was shown to be large enough to exhibit short path lengths, and at the same time small enough to exhibit a high clustering coefficient of the network [4]. In this small world setup every node has the same number of subscriptions.

*3) Scale-free networks*

The first well-established work on scale-free networks was done by Barabási and Albert. In scale-free networks, the degree of the nodes is distributed by power-law. To generate random undirected scale-free networks, Barabási and Albert have proposed a simple algorithm: The algorithm starts with an initial (small) network that is grown to a scale-free network. Every new node that is added to the network in the growing phase has exactly *k* edges, which are linked to the existing network by using preferential attachment. This means, the more edges an existing vertex has, the more likely the new vertex will connect to it. Unfortunately, the Barabási-Albert model cannot be directly transferred into a directed network. There has been a lot of work on scale-free networks and also several attempts at modelling a directed version. The various attempts mostly differ in how the initial small network that is to be grown is first generated, and in how the problem that directed scale-free networks are often acyclic, unlike real world networks is solved [12].

For our work, the implementation proposed by Yuan and Wang [14] is used. This starts with a fully connected network, and then grows it using preferential attachment. This results in a citation-network with feed-forward characteristics, where each new node is connected pointing towards the existing nodes. However, with a small probability *p* the direction of edges is inverted, to allow the generation of cycles. We used *p*=0.15, which was a value under which Yuan and Wang could find the network to be "at the edge of chaos" when transformed into Kauffman's NK model [15].

Here, *k* is the average degree of the final network, and thus the number of edges introduced to the existing network with every new vertex added. The starting network is a set of *k* fully connected vertices. In the growing phase, from the *k+1*th vertex onwards, the algorithm connects the current node to or from the initial set using preferential attachment. Thus, in the scale-free version the number of subscriptions each node has is not the same, but distributed by a power law.

## V. RESULTS

Here we discuss the outcome of SPECI simulations where the networks of subscriptions, which are being polled for aliveness, have each of the structures presented in Section IV. These new results are then contrasted with the initial results from SPECI, as were shown in Section III.

This section shows the impact of the distribution of the subscriptions on the scaling properties of the protocol. It is further shown that the effect of the subscription topology that emerges under the Transitive P2P protocol cannot be observed under the other protocols, as shown in the example of the hierarchical protocol, here.

### A. Transitive P2P protocol

First, we will analyse the impact of the subscription network on a data centre with a Transitive P2P protocol. Figure 5 and Figure 6 show the average percentages of nodes that have at least one subscription, which is inconsistent with the real state. Figure 5 shows this on logarithmic scale and Figure 6 on a normal scale. In Figure 5 it can be seen, that for large *n* and for high failure rates there is a direct relation between the clustering coefficient of the subscription graph and the number of inconsistencies. For $n=10^4$ and failure rate of 10%, for example, Figure 6 illustrates that there are statistically significantly fewer inconsistencies under randomly distributed subscriptions than under Barabási-Albert scale-free distributed subscriptions; fewer under scale-free than under Watts-Strogatz small-world subscription networks; and fewer under small-world than under subscriptions distributed on a regular lattice or grid. In fact, in Figure 6 it can be seen that for $n=10^4$ and failure rates of *f*=1% and *f*=10%, and for subscription distributions with a high clustering coefficient (i.e. lattice and small-world) the rate even reaches and exceeds a mean of 50% of inconsistencies. This suggests that with such *f* and *n* the system will not return to a consistent state. The growth of inconsistencies slows down when even higher failure rates are reached, because the likelihood increases that a newly introduced change or failure has no impact (when all subscribed nodes are already inconsistent) or even improves the state (when the change reverts inconsistent views to become consistent). Growth to a mean above 50% is however possible, because each node has multiple subscriptions, but counts as inconsistent as soon as one of the subscriptions is inconsistent. The effect of this increase in inconsistencies emerges when the subscription network forms cycles, which at a high frequency pass on invalid data, before it gets outdated or recognised as incorrect.

Figure 7 shows the average load generated on every node when the Transitive P2P protocol is used with different types of subscriptions. When the subscriptions have a low clustering coefficient, the load grows with the increasing $n$. In contrast, when the subscriptions are distributed with a high clustering coefficient, there is little growth of the load on the nodes. Because of the high clustering, in the Transitive P2P protocol a node can update a large fraction of its subscriptions with a single access. However, it becomes necessary to counteract the sharp rise in inconsistencies by tuning the middleware. This can be done, but only at the cost of additional load.

This example shows that the placement and structure of components that collaborate can have an impact on the performance of the middleware, and needs to be taken into account when tuning and selecting the middleware topology. For instance, if it is known that the subscriptions will have a distribution with a high clustering coefficient, it is desirable to make up for inconsistencies by increasing the polling interval at the cost of additional load. It also demonstrates the need for rigorous simulation tools to allow planning the design of large-scale data centres and to avoid complex effects to occur that were not anticipated or accounted for at design phase.

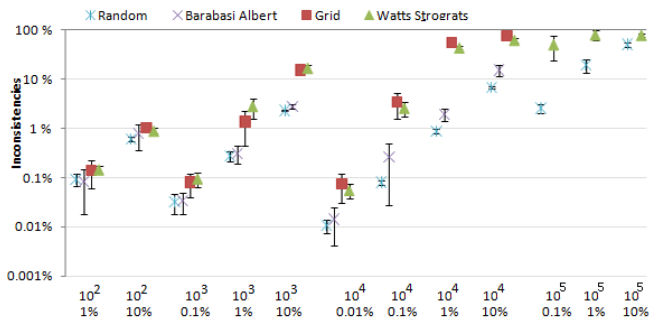

Figure 5. Inconsistencies under varying subscription types, logarithmic scale

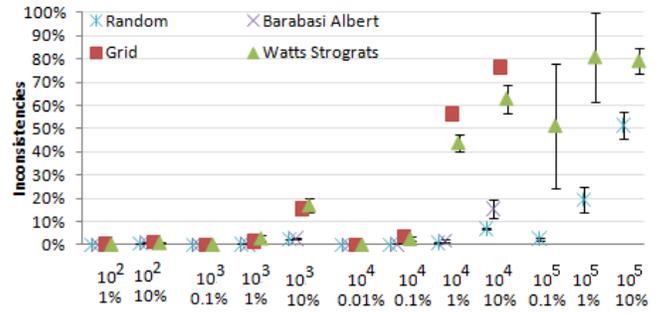

Figure 6. Inconsistencies under varying subscription types, linear scale

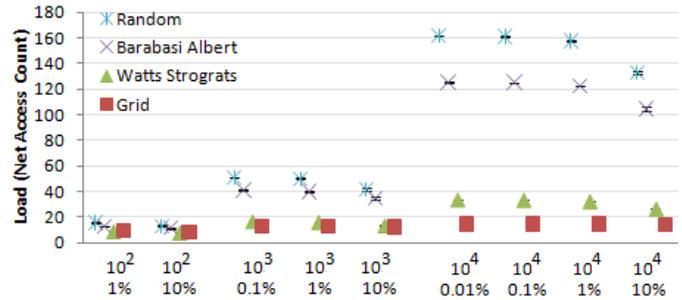

Figure 7. Load of Transitive P2P with different distributions of subscriptions

## B. Scaling properties of the subscription topologies:

In this subsection we compare the scaling properties the Transitive P2P protocol exhibits when the subscription network has scale-free characteristics, with the case when the subscription network has small-world characteristics. The mean of inconsistencies when the failure rate is 1% and 10% is plotted in Figure 8. For $n = 10^2$, there is no significant difference between the two subscription distribution. With increasing $n$, in the small-world case the number of inconsistencies grows a lot faster than in the scale-free setting. Figure 9 shows the average load generated in the same case. Here, the small-world topology exhibits the benefits of the high clustering, which allows a node to update a large fraction of its subscriptions with few net accesses. With increasing $n$, in the scale-free setting the load grows a lot faster than in the small-world setting.

For illustration, in Figure 8 and Figure 9 a straight line along the small-world and the scale-free values has been added. At first sight, the difference in slope seems to be larger in the case of inconsistencies than in the case of the load.

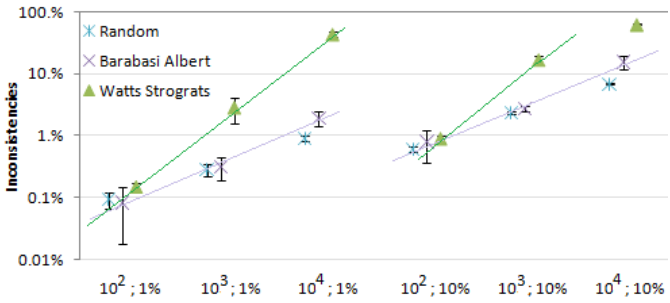

Figure 8. Inconsistencies grouped by failure rate

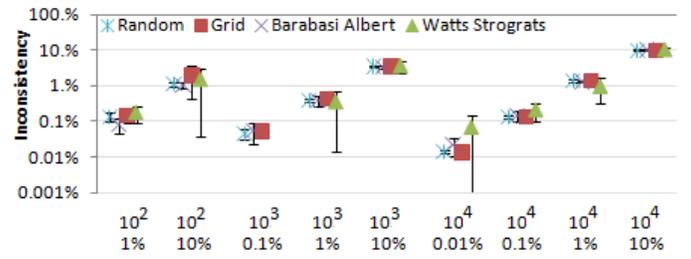

Figure 10. Inconsistencies under hierarchical topology

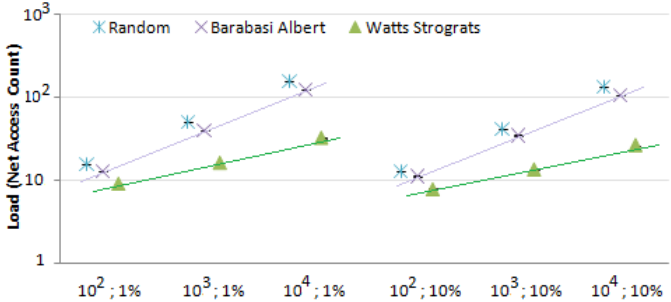

Figure 9. Load grouped by failure rate

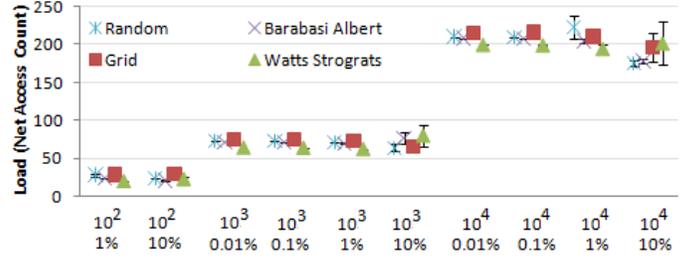

Figure 11. Load under hierarchical topology

*C. Hierarchical protocol*

The previous subsections showed that the topology of subscriptions plays a significant role when assessing the behaviour of the Transitive P2P protocol. This subsection shows that under the hierarchical protocol the topology of subscriptions plays no significant role for inconsistencies or load.

Figure 10 shows the mean of inconsistencies and their confidence intervals on logarithmic scale. For each *n* and failure rate, the outcome of the simulations for the four subscription topologies is plotted. The simulations returned no statistically significant difference between the runs with different subscription topologies. Figure 11 shows the average load and its confidence intervals for the same runs displayed in Figure 10. Again, no significant difference could be seen between the runs. Both graphs exhibit the same properties as already discussed in the baseline SPECI results (Section III).

The varying clustering coefficient in the distribution of the subscriptions, which leads to differences in the behaviour of the transitive P2P protocol, plays no role in the hierarchical protocol. Further, the power-law distribution of the number of subscriptions in the scale-free topology (i.e. that some nodes have only few subscriptions while others have a far larger number of subscriptions), as opposed to the other distributions investigated where every node has the same number of subscriptions, does not appear to have an impact on the number of inconsistencies or the average load in the hierarchical protocol, either.

*D. Discussion of the findings*

This section has shown the outcome of simulations where the topology of the networks of subscriptions, which are being polled for aliveness, have various structures.

The key finding is that there is a direct relation between the clustering coefficient of the subscription graph and the number of inconsistencies found when using scalable transitive P2P protocols. This means that not only the number of components and the number of subscriptions are relevant to determine the scaling properties of data centre middleware, but also the topology of the network formed by the distribution of the subscriptions. It also means component placement has a direct effect on how the data centre performance scales. Based on our findings, we suggest that when designing and tuning data centre middleware, it is important to analyse how the components are placed, and how they interact: not just in terms of number but also in terms of the structure of their interaction. In this way, depending on the form of subscriptions, scalable P2P middleware can be tuned to use either those polling intervals or those maximum allowed ages of data to be passed on that result in the desired Quality of Service for the data centre.

In Section III we raised the issue that for the Transitive P2P protocol we would have expected a larger number of inconsistencies than we observed. The results in this section have given the answer to this question. When the underlying distribution of subscriptions is a random graph with a low clustering coefficient, the Transitive P2P protocol behaves differently to when the underlying subscriptions have a high clustering coefficient. In the latter case there are many common subscriptions, and data is passed on transitively. This reduces the load, but at the same time increases the risk that data passed on accumulates delays that are not noticed, leading to a high number of inconsistencies.

At the same time, it has been shown that the structure of the subscription topology, in contrast to the case with the Transitive P2P protocol, did not have any impact on the hierarchical protocol.

Finally, our results emphasise the need for rigorous simulation tools such as SPECI to allow planning the design of large-scale data centres, as well as to explore and avoid complex effects, which were perhaps not expected or taken into consideration at the design phase.

## VI. FUTURE WORK

We plan further case studies with the SPECI simulator, exploring alternative models for the problems described, and expanding the simulator. There are four case study scenarios of immediate interest. First, the relationship between the load results and the inconsistency results, such as the issue of at which loads each of the protocols can perform within different maximum tolerable levels of inconsistencies. Second, special cases of spatially correlated failure conditions (where a large number of physically adjacent nodes fail simultaneously, such as when a DC aisle loses its provision of cooled air-conditioning). Along with this, one can simulate various recovery mechanisms, where failed components are not replaced until the majority of components in a rack have failed and then the entire rack gets replaced. Third, it needs to be explored what combined rates of failure, number of subscriptions, and load thresholds prevent the system from ever reaching a consistent status, and what settings make it impossible for the system to recover from correlated failure. Fourth, instead of assuming uniform costs for connecting to other nodes, we plan to explore how varying the connection cost affects component placement recommendations.

For medium-term future work, we intend to look for alternative models to verify the findings from the simulator. Formal analytical mathematical models have some appeal, but we also intend to verify our SPECI simulator outputs against real-world DCs, where that is possible. Finally, SPECI needs to be expanded. At the moment we have used it only to explore the one-dimensional state communication problems discussed here. Real middleware has to distribute several policies over the network. It needs to account for virtual machines and load-balancing, security, and job scheduling. There is a need for rigorous simulation tools that are capable of modelling such multidimensional problems that middleware is facing in order to access the design and scaling properties of future ultra-large scale DCs.

While we are acutely aware that SPECI can be improved and extended in many ways, we feel that the results presented here demonstrated the worth of our approach. The use of advanced computer simulation techniques has become commonplace in many areas of science and engineering (see, e.g., [16, 17] for discussion). Recent work has explored the use of multi-agent simulation techniques for engineering analysis of large-scale systems-of-systems [18], and has presented some simple minimal models for exploring various aspects of cloud computing [19, 20, 21]. Nevertheless, to the best of our knowledge there are currently no simulation tools for cloud-scale data-centres that operate at the level of analysis that SPECI is aimed at serving.

## VII. CONCLUSION

In the field of engineering of complex computer systems, despite the well-established tradition of computational modelling and simulation tools in other engineering domains, there are currently no comparable tools for cloud-scale computing data-centres. The lack of such tools prevents the application of rigorous formal methods for testing and verifying designs before they go into production.

As an exploratory step in meeting this need, we have developed SPECI. The initial version of SPECI has been used to explore the simple case where within the data-centre there are components that work together and need to know the status of the other components via "subscriptions" to status-updates from those components.

In [1] we used a first-approximation assumption that such subscriptions are distributed wholly at random across the data centre.

In this present paper, we explored the effects of introducing more realistic constraints to the structure of the internal network of subscriptions. We contrasted the original results from SPECI with new results from simulations exploring the effects of making the data-centre's subscription network have a regular lattice-like structure, and also semi-random network structures resulting from parameterised network generation functions that create small-world and scale-free networks.

We showed that when using scalable transitive P2P middleware there is a direct relation between the clustering coefficient of the graph that describes the structure of the subscription network and the number of inconsistencies found. This means, that not only the number of components and the number of subscriptions are relevant to determine the scaling properties of data centre middleware, but also the distribution of the subscriptions and the nature of their interactions. It also means component placement has a direct effect on how the data centre performance scales. Based on our findings we suggest that when designing and tuning data centre middleware, it is important to analyse how the components are placed and how they interact, not just in terms of their number but also in terms of the structure of their interaction. In this way, scalable P2P data centre middleware can be tuned, depending on the form of subscriptions, to use such polling intervals or maximum allowed "time to live" ages for data to be passed on, that result in the desired quality of service for the data centre.

The results presented in Section V showed the potential value of rigorous simulation tools. There is scope for much further work well beyond the current state of SPECI, to allow planning the design of large-scale data centres, and in order to explore and avoid complex emergent effects that were not accounted for at design phase

We see the results in this paper as the first step towards developing adaptive data-centre management policies that "intelligently" and dynamically organise and reorganise the network of subscriptions within the data-centre in the light of changing demands, while best ameliorating the effects of normal failure. SPECI is a first step; there is manifestly much more to be done.

ACKNOWLEDGEMENTS

This work is funded by Hewlett-Packard Labs: in addition to their financial sponsorship, we are very grateful to John Manley, Patrick Goldsack, & Johannes Kirschnick at HP for valuable conversations. We are also grateful to our colleagues in the UK Large-Scale Complex IT Systems (LSCITS) research programme, Ali Khajeh-Hosseini, John Rooksby, & Ian Sommerville, for conversations and comments; and to EPSRC for their funding of the LSCITS Initiative.

REFERENCES

[1] I. Sriram, "SPECI, a simulation tool exploring cloud-scale data centres," in CloudCom 2009, LNCS 5931, pp. 381-392, M.G. Jaatun, G. Zhao, and C. Rong (Eds.), Springer-Verlag Berlin Heidelberg 2009 http://arxiv.org/abs/0910.4568

[2] L Barroso and U. Hölzle, "The datacenter as a computer: An introduction to the design of warehouse-scale machines," Synthesis Lectures on Computer Architecture Volume 4, 2009

[3] L. W. Nagel, "SPICE2: A computer program to simulate semiconductor circuits," Technical Report No. ERL-M520, University of California, Berkeley 1975

[4] D. Watts and S. Strogatz, "Collective dynamics of 'small-world' networks," *Nature* 393, 4th June 1998, pp. 440-442

[5] A.-L. Barabási and R. Albert, "Emergence of scaling in random network," *Science*, 286, 1999, pp. 509-512

[6] R. Albert and A.-L. Barabási, "Statistical Mechanics of complex networks," *Rev. Mod. Phys.* vol. 74:47, 2002.

[7] Jogalekar, P.; Woodside, M., "Evaluating the scalability of distributed systems," *Parallel and Distributed Systems, IEEE Transactions on*, vol.11, no.6, pp.589-603, Jun 2000

[8] M. Sloman, "Policy driven management for distributed systems," Journal of Network and Systems Management, Vol. 2 no.(4), pp. 333 - 360, 1994-12-30

[9] R Ranjan and R Buyya, "Decentralized overlay for federation of Enterprise Clouds", In K-C Li et al., Handbook of Research on Scalable Computing Technologies, IGI Global, 2010, pp. 191-218.

[10] Kempe, D., Dobra, A., and Gehrke, J. 2003. Gossip-Based Computation of Aggregate Information. In *Proceedings of the 44th Annual IEEE Symposium on Foundations of Computer Science* (October 11 - 14, 2003). FOCS. IEEE Computer Society, Washington, DC, 482.

[11] K. Birman, "The promise, and limitations, of gossip protocols," SIGOPS Oper. Syst. Rev. 41, 5, Oct 2007, pp. 8-13.

[12] M. E. J. Newman, "The Structure and Function of Complex Networks," *SIAM Review*, 45:167-256, http://arxiv.org/abs/cond-mat/0303516, 2003

[13] J. Kleinberg. "The small-world phenomenon: An algorithmic perspective," *Proc. 32nd ACM Symp. on Theory of Computing*, 2000.

[14] B. Yuan, and B. Wang, "Growing directed networks: organisation and dynamics," *New J. Physics*, 9(8), pp. 282-290, http://arxiv.org/pdf/cond-mat/0408391, 2007.

[15] S. Kauffman, *"The Origins of Order,"* OUP, 1993.

[16] S. Turkle, *"Simulation and its Discontents,"* MIT Press, 2009.

[17] T. Hey, et al. (eds). *"The Fourth Paradigm: Data-Intensive Scientific Discovery,"* Free e-book available at http://research.microsoft.com/en-us/collaboration/fourthparadigm/, 2009.

[18] R. Alexander, *"Using Simulation for System of System Hazard Analysis,"* PhD Thesis, Dept. of Computer Science, University of York, 2007.

[19] J. Weinman (2009), website at http://www.complexmodels.com

[20] R. Buyya, R. Ranjan, and R. N. Calheiros, "Modeling and Simulation of Scalable Cloud Computing Environments and the CloudSim Toolkit: Challenges and Opportunities," In: Proceedings of the 7th High Performance Computing and Simulation (HPCS 2009) Conference, Leibzig, Germany, 2009

[21] D. G. Murray and S. Hand, "Nephology: Measuring heterogeneity in cloud computing", poster presented at NSDI 2009, Boston http://www.cl.cam.ac.uk/~dgm36/publications/2009-nephology-nsdi-poster.pdf, 2009